\documentclass[proof]{pasj00}
\draft

\begin{document}
\SetRunningHead{Y. Shioya et al.}{SDSSp J104433.04-012502.2 at $z=5.74$}
\Received{2002/06/18}
\Accepted{2002/10/6}

\title{SDSSp J104433.04$-$012502.2 at $z=5.74$ is Gravitationally Magnified
       by an Intervening Galaxy\altaffilmark{*}}

\author{Yasuhiro \textsc{Shioya}      \altaffilmark{1},
        Yoshiaki  \textsc{Taniguchi}  \altaffilmark{1},
        Takashi  \textsc{Murayama}    \altaffilmark{1},
        Masaru  \textsc{Ajiki}        \altaffilmark{1},\\
        Tohru  \textsc{Nagao}         \altaffilmark{1},
        Shinobu S.  \textsc{Fujita}   \altaffilmark{1},
        Yuko  \textsc{Kakazu}         \altaffilmark{2},
        Yutaka  \textsc{Komiyama}     \altaffilmark{3},\\
        Sadanori  \textsc{Okamura}    \altaffilmark{4,5},
        Shinki  \textsc{Oyabu}        \altaffilmark{6},
        Kimiaki  \textsc{Kawara}      \altaffilmark{6},
        Youichi  \textsc{Ohyama}      \altaffilmark{3},\\
        Koji S.  \textsc{Kawabata}    \altaffilmark{7},
	Hiroyasu  \textsc{Ando}       \altaffilmark{3},
        Tetsuo  \textsc{Nishimura}    \altaffilmark{3},
        Masahiko  \textsc{Hayashi}    \altaffilmark{3},\\
        Ryusuke  \textsc{Ogasawara}   \altaffilmark{3}, \&
        Shin-ichi  \textsc{Ichikawa}  \altaffilmark{7}
	}

\altaffiltext{1}{Astronomical Institute, Graduate School of Science,
        Tohoku University, Aramaki, Aoba, \\ Sendai 980-8578}
\email{shioya@astr.tohoku.ac.jp}
\altaffiltext{2}{Institute for Astronomy, University of Hawaii,
        2680 Woodlawn Drive, Honolulu, HI 96822, USA}
\altaffiltext{3}{Subaru Telescope, National Astronomical Observatory of Japan,
        650 N.A'ohoku Place, Hilo, HI 96720, USA}
\altaffiltext{4}{Department of Astronomy, Graduate School of Science,
        The University of Tokyo, Tokyo 113-0033}
\altaffiltext{5}{Research Center for the Early Universe, School of Science,
        The University of Tokyo, Tokyo 113-0033}
\altaffiltext{6}{Institute of Astronomy, Graduate School of Science,
        The University of Tokyo, 2-21-1 Osawa, Mitaka, \\ Tokyo 181-0015}
\altaffiltext{7}{National Astronomical Observatory,
        2-21-1 Osawa, Mitaka, Tokyo 181-8588}


\maketitle

\footnotetext[*]{Based on data collected at 
	Subaru Telescope, which is operated by 
	the National Astronomical Observatory of Japan.}

\begin{abstract}
During the course of our optical deep survey program on
L$\alpha$ emitters at $z \approx 5.7$ in the sky area
surrounding the quasar SDSSp J104433.04$-$012502.2
at $z=5.74$, we found that a faint galaxy with
$m_B$(AB) $\approx 25$ is located at \timeform{1".9} southwest of the quasar.
Its broad-band color properties from $B$ to $z^\prime$ suggest that 
the galaxy is located at a redshift of $z \sim 1.5$ --- 2.5. 
This is consistent with
no strong emission line in our optical spectroscopy. Since the counter
image of the quasar cannot be seen in our deep optical images, the 
magnification factor seems not to be very high. Our modest estimate is that 
this quasar is gravitationally magnified by a factor of 2.
\end{abstract}

\section{Introduction}

The Sloan Digital Sky Survey (SDSS: e.g., York et al. 2000)
has been finding high-redshift quasars at $z \approx 6$; 
the most distant known to date is SDSSp J103027.10+052455.0
at $z=6.28$ (Fan et al. 2000, 2001).
One serious problem seems to be that all of these $z \sim 6$ SDSS quasars 
are exceptionally bright (see table 1). 
As claimed by Wyithe and Loeb (2002; hereafter WL02),
the presence of such very bright quasars at $z \sim 6$ raises
the following two problems: (1) If the central engine of these
quasars is an accreting supermassive black hole and shines
near the Eddington accretion rate, a supermassive black hole 
with mass exceeding $\sim 3 \times 10^9 M_\odot$ was already formed
beyond $z=6$. This challenges models for early structure formation
(Turner 1991; Haiman, Loeb 2001). Also, (2) the absolute luminosities
of the $z \sim 6$ quasars are systematically higher than
expected from the SDSS survey criteria (Fan et al. 2000, 2001),
suggesting that the luminosity function of the $z \sim 6$ quasars
is significantly biased to higher luminosities. 
These two problems urge WL02 to propose that the SDSS $z \sim 6$ quasars
may be magnified by intervening gravitational lenses.

During the course of our optical deep survey program on
L$\alpha$ emitters at $z \approx 5.7$ in the field surrounding
the quasar SDSSp J104433.04$-$012502.2 at $z=5.74$,
we found that
a faint galaxy with $m_B$(AB) $\approx 25$ is located at
\timeform{1".9} southwest of this quasar.
In this paper, we report on the possibility that 
SDSSp J104433.04$-$012502.2 is gravitationally magnified
by a faint galaxy with optical photometric and 
spectroscopic data obtained by our observations.
Throughout this paper, 
we adopt a flat universe with $\Omega_{\rm m} = 0.3$,
$\Omega_{\Lambda} = 0.7$, 
and $h=0.7$, where $h = H_0/($100 km s$^{-1}$ Mpc$^{-1}$).

\section{Optical Deep Imaging}

Deep optical imaging observations were made with the Suprime-Cam
(Miyazaki et al. 1998) on 
the 8.2 m Subaru telescope (Kaifu 1998) at Mauna Kea Observatories.
The Suprime-Cam consists of ten 2k$\times$4k CCD chips and 
provides a very wide field of view: $34^\prime \times 27^\prime$ 
with a \timeform{0".2}/pixel resolution.
Therefore, the combination between the Suprime-Cam and the 
Subaru telescope enables us to carry out
wide and deep narrow-band imaging surveys for high-$z$ emission-line objects.
Using this facility, we made a very deep optical imaging survey for
faint L$\alpha$ emitters in the field surrounding the SDSSp
J104433.04$-$012502.2 at a redshift of 5.74 (Fan et al. 2000;
Djorgovski et al. 2001; Goodrich et al. 2001). 
In this survey, we used a narrow-passband filter, NB816, centered at
8160 \AA ~ with a passband of $\Delta\lambda$(FWHM) = 120 \AA;
the central wavelength corresponds to a redshift of 5.72 for 
L$\alpha$ emission. 
We also used broad-passband filters: 
$B$, $R_{\rm C}$, $I_{\rm C}$, and z$^{\prime}$. A summary of the imaging observations
is given in table 2. All of the observations were made under
photometric conditions, and the seeing size was between \timeform{0".7}
and \timeform{1".3} during the run.

In this paper, we analyze only two CCD chips in which
the quasar SDSSp J104433.04$-$012502.2 is included
(see also Ajiki et al. 2002).
The field size is \timeform{11'.67} by \timeform{11'.67}.
The individual CCD data were reduced and
combined using IRAF and mosaic-CCD data-reduction software
developed by Yagi et al. (2002).
The photometric and spectrophotometric
standard stars used in the flux calibration were SA101 for the
$B$, $R_{\rm C}$, and $I_{\rm C}$ data, and GD 50, GD 108
(Oke 1990), and PG 1034+001 (Massey et al. 1996)
for the NB816 data. The $z^\prime$ data were
calibrated using the magnitude of SDSSp J104433.04$-$012502.2
(Fan et al. 2000).
Thumb-nail optical images of the quasar filed are shown in figure 1.
A faint galaxy is seen at \timeform{1".9} southwest from the quasar.
This galaxy is seen even in the $B$-band image, giving a rough 
constraint such that its redshift is $z < 2.6$, 
since the wavelength of redshifted L$\alpha$ absorptions by 
intergalactic neutral hydrogen is considered to be shorter 
than about 4400 \AA .

\section{Optical Spectroscopy}

In order to estimate a spectroscopic redshift,
we carried out optical
spectroscopy using the Subaru Faint Object Camera And Spectrograph
(FOCAS; Kashikawa et al. 2000)  on 2002 March 13 (UT).
We used 300 lines mm$^{-1}$ grating blazed at 7500 \AA~ (300R) together
with an order cut filter SY47, giving a spectral resolution of 
$R \simeq 1000$ with a \timeform{0".8}-wide slit.
This setting gave a  
wavelength coverage between 4700 \AA ~ and 9400 \AA. 
The integration time was 1800 s. 
The spectrum is shown in figure 2. 
This spectrum shows that this galaxy has blue colors, and 
that there is no prominent emission line.
If this galaxy is a strong emission-line galaxy, like a 
starburst galaxy, it is inferred either that 
its redshifted wavelength of L$\alpha$ emission line
would be shorter than 4700 \AA,  ~ or that its redshifted wavelength of
[O {\sc ii}]$\lambda$3727 emission line would be longer than
9400 \AA. From these constraints, one may estimate that the redshift of
this galaxy lies in a range between $z \approx 1.5$ and $z \approx 2.9$.

\section{Results and Discussion}

The faint galaxy which we have found (hereafter, the lens galaxy)
is located at \timeform{1".9} southwest 
from the quasar. In order to estimate how the quasar is 
gravitationally magnified, we need information about both the redshift
and the stellar velocity dispersion of this lens galaxy. However, 
no direct information on these two quantities was obtained in our 
observations.
Therefore, in this paper, first, we estimate its photometric redshift 
based on our optical broad-band photometric data (subsection 4.1).
Second, we also estimate its stellar velocity dispersion 
using the Tully---Fisher relation (subsection 4.2). Then, we estimate
the magnification factor based on the singular isothermal sphere model
for gravitational lensing (subsection 4.3).

\subsection{Photometric Redshift of the Lens Galaxy}

Using optical $B$, $R_{\rm C}$, $I_{\rm C}$, NB816, and $z^\prime$
data, we estimated a probable photometric redshift of the lens galaxy.
Its optical magnitudes obtained in our observations were 
$B = 25.1 \pm 0.06$, $R_{\rm C}  = 24.5 \pm 0.04$, $I_{\rm C} = 24.3 \pm 0.06$, 
${\rm NB}816 = 24.2 \pm 0.04$, and $z^\prime = 23.9 \pm 0.09$; the quoted errors
are 1$\sigma$ sky noises. 
It is noted again that because the lens galaxy is not a $B$-dropout, 
we had a
constraint of $z < 2.62$.

A photometric redshift of a galaxy was evaluated from the global shape of
the spectral energy distribution (SED) of a galaxy,
as a redshift with the maximum likelihood,

\begin{equation}
L(z,t)=\prod_{i=1}^{5} \exp \left\{ - \frac{1}{2} 
\left[ \frac{f_i-AF_i(z,t)}{\sigma_i} \right]^
2 \right\},
\end{equation}
where $f_i$, $F_i(z, t)$, and $\sigma_i$  are the observed flux, 
the template flux, and the error of the $i$-th band, respectively, 
and $A$ is defined as

\begin{equation}
A = \frac{\sum F_i f_i/\sigma_i^2}{\sum F_i^2/\sigma_i^2}.
\end{equation}

The SED of a galaxy that we observed was
mainly determined by the following four factors:
(1) the radiation from stars in the galaxy,
(2) the extinction by dust in the galaxy, itself,
(3) the redshift of the galaxy, and 
(4) the absorption by the intergalactic neutral hydrogen between
the galaxy and us.
We treated the above factors in the following way.

We used the population synthesis model GISSEL96,
which is a revised version of Bruzual and Charlot (1993)
(see also Leitherer et al. 1996), to calculate an SED from stellar components.
The SED of the galaxy was determined by its star-formation history.
The SEDs of local galaxies were well reproduced
by models whose star-formation rate
declines exponentially ($\tau$ model); i.e., 
$SFR(t) \propto \exp (-t/\tau)$, 
where $t$ is the age of the galaxy and $\tau$ is the time scale of star formation.
Different combinations of $t$ and $\tau$ generate a similar shape of SED. 
In this work, 
we therefore used $\tau=1$ Gyr models with the Salpeter's initial mass function
(the power index of $x=1.35$ and the stellar mass range of 
$0.01 \leq m/M_\odot \leq 125$) to derive various SED types.
We adopted the solar metallicity, $Z=0.02$. 
For these models, we calculated SEDs with ages of
$t$ = 0.1, 0.5, 1, 2, 3, 4, and 8 Gyr, and 
called them SED1, SED2, SED3, SED4, SED5, SED6, and SED7, respectively; 
note that the SED templates derived by Coleman et al. (1980), elliptical 
galaxies (the bulges of M 31 and M 81), Sbc, Scd, and Irr
correspond to those of SED7 ($t=8$ Gyr), SED6 ($t=4$ Gyr), SED5 ($t=3$ Gyr), 
and SED4 ($t=2$ Gyr), respectively.
We adopted the dust extinction curve for starburst galaxies determined 
by Calzetti et al. (2000) with visual extinctions of 
$A_V$ = 0, 0.2, 0.4, 0.6, 0.8, and 1.0.
As for the absorption by intergalactic neutral hydrogen,
we used the average optical depth derived by Madau et al. (1996).
However, Scott, Bechtold, and Dobrzycki (2000) showed that 
the observed continuum depression between $1050 \; (1+z)$ \AA~ and $1170
\; (1+z)$ \AA~
due to the L$\alpha$ clouds ($D_{\rm A}$) for quasars with $z \lesssim$ 3
is lower than that expected from extrapolation using the 
cosmic transmission derived by Madau et al. (1996).
Further, they also showed that there is a scatter in $D_{\rm A}$ from 
quasar to quasar. Therefore, we also investigated the following two
cases: 1) 0.5 $\times$ the Madau et al.'s average optical depth,
and 2) 2 $\times$ the Madau et al.'s average optical depth.
We then estimated the probable photometric redshift of the lens galaxy.
In this procedure, we adopted an allowed redshift of between $z=0$ and 
$z=6$ with a redshift bin of $\Delta z$ = 0.02.

Our results are shown in figure 3. 
In this figure, we show the distributions
of the likelihood for the seven SEDs.
Our best estimate of the photometric redshift of the lens galaxy
is $z_{\rm phot} \approx 2.1$. Since the likelihood, $L(z,T)$, exceeds
0.1 for $z \sim$ 1.5 --- 2.5, it is suggested that the lens galaxy
is located at 1.5 $< z <$ 2.5. This range is consistent with 
the information from our optical spectroscopy (section 3). 
SED1 appears to be inappropriate for the SED of the lens galaxy.
It is found that SED7 also has a maximum likelihood larger than 0.1 
around $z \sim 3$. 
However, since the age of the galaxy at $z \sim 3$ is considered to be 
younger than 2 Gyr, we also threw out SED7.
The luminosity of the galaxy was evaluated to be   
the normalization factor, $A$, times the luminosity of the model. 
The range of the evaluated absolute $B$-band magnitude 
for SED2 through SED6 was between $z=1.5$ and 2.5 is $-20.4 > M_B > -23.1$.

\subsection{Stellar Velocity Dispersion of the Lens Galaxy}

Although it is necessary for the estimate of the magnification factor
by gravitational lensing to know the stellar velocity dispersion
($\sigma_v$) of 
the lens galaxy, there is no direct measurement.
Given the redshift of a lens galaxy, we could estimate its absolute
magnitude and then estimate a probable value of the stellar velocity
dispersion. Since it is considered that the lens galaxy is located at
1.5 $< z <$ 2.5, we could estimate the stellar velocity dispersion 
as a function of both the redshift and SED based on the Tully---Fisher relation 
established for a sample of galaxies in the local universe. 

Here, we used the Tully---Fisher relation derived by Sakai et al. (2000),

\begin{equation}
M_B  =  -8.07 (\log W_{20} - 2.5) - 19.88,
\end{equation}
where $M_B$ is the absolute $B$ magnitude and
$W_{20}$ is the full width at 20\% of the maximum velocity.
Assuming that $\sigma_v = V_{\rm rot}/\sqrt{2} = W_{20}/2\sqrt{2}$ where
$V_{\rm rot}$ is the rotation velocity, 
we estimated $\sigma_v$ as a function of $z$ and SED.
We evaluated $M_B$ using the method given in subsection 4.1.
The results are shown in figure 4. Since the cases of SED1 and SED7 are 
rejected in subsection 4.1, the velocity dispersion is estimated 
to be in a range between 140 km s$^{-1}$ and 280 km s$^{-1}$ for a
redshift range of 1.5 $< z <$ 2.5.

\subsection{Magnification Factor by the Gravitational Lensing}

We are now ready to estimate the magnification factor by 
gravitational lensing. We adopt the singular isothermal sphere
(SIS) model for simplicity (e.g., Binney, Merrifield 1998).
In this model, the magnification factor for the brighter source 
can be expressed as

\begin{equation}
M_+ = \frac{\theta}{\theta - \theta_{\rm E}},
\end{equation}
where $\theta$ is the angle between the lens galaxy and 
the source (i.e., SDSSp J104433.04$-$012502.2 in this case),
and the $\theta_{\rm E}$ is the Einstein angle, defined as

\begin{equation}
\theta_{\rm E} = 4 \pi \left( \frac{\sigma_v}{c} \right)^2 
\frac{D_{\rm LS}}{D_{\rm OS}},
\end{equation}
where $D_{\rm LS}$ is the angular diameter distance between the
lens galaxy and the source and $D_{\rm OS}$ is that between
the observer and the source. In figure 5, we show 
the magnification factor as a function of $z$  
for eight cases
of $\sigma_v$ = 140, 160, 180, 200, 220, 240, 260, and 280 km s$^{-1}$.
In this figure, we also show the magnification factor for
the counter image [$M_- = \theta/(\theta_E - \theta)$], 
which can be seen at $\theta = \theta_{\rm E} - \beta$,
where $\beta$ is the angle between the lens galaxy and the intrinsic source.
If the lens galaxy is located at $z \sim 2$ and has $\sigma_v \sim
280$ km s$^{-1}$, $\theta_{\rm E}$ becomes \timeform{1".9} and we thus
obtain $M_+$ = infinity. However, if this is the case, 
since $\beta \sim 0$ and $M_- \sim M_+$, we would observe 
the counter image with nearly the same brightness as that of
the lensed image. 
Further, if $M_+ > 10$, $M_-$ is also as large as $M_+$. 
As we shown in figure 1 that there is no apparent object, 
except the quasar ($z^{\prime} = 19.2$ mag) and 
the lens galaxy ($z^{\prime} = 23.9$). 
Since there is no apparent counter image, the magnification
factor must be much less than 10. 
If $\sigma_v \lesssim$ 200 km s$^{-1}$, the condition of 
$M_- \ll M_+$ is achieved at $z \sim 2$. In this case, 
we obtain $M_+ \sim 2$. 

\subsection{Concluding Remarks}

We have found a candidate of a lensing galaxy at \timeform{1".9} 
southwest of the SDSS high-$z$ quasar, SDSSp J104433.04$-$012502.2.
From our observational data, we find that the lens galaxy
is located at $1.5 < z < 2.5$; the most probable photometric
redshift is $z_{\rm phot} \approx 2.1$. Adopting the SIS model
for gravitational lensing, we estimate the magnification factor to be 
$M_+ \sim 2$ if the lens galaxy is located at $z \sim 2$ and has
a stellar velocity dispersion of $\sigma_v \sim 200$ km s$^{-1}$.

WL02 suggested that the SDSS $z \sim6$ quasars are systematically brighter
than expected. Namely, they investigated the probability that
a quasar with a rest-frame 1450 \AA ~ magnitude, $M_{1450}$(AB), and
a redshift of $z$ was selected into the survey using the survey selection 
function given in Fan et al. (2001). They found that 
the cumulative probability of $M_{1450}$(AB) for the $z \sim 6$ quasars
is different from the expected one at the 95\% significance level.
However, they also suggested that this inconsistence is removed if 
the flux of one or more of the $z \sim 6$ quasars is overestimated by a 
factor of 2. Therefore, our finding presented in this paper may
remove the inconsistency raised by WL02.

It is known that the lensing optical depth increases with increasing
the redshift, and thus objects at higher redshift are more affected
by gravitational lensing from a statistical point of view
(e.g., Turner 1991; Barkana, Loeb 2000; WL02). As proposed by WL02,
it will be very important to carefully search for lensing galaxies
towards all high-$z$ quasars by deep and high-resolution
imaging.

\bigskip

We would like to thank the Subaru Telescope staff
for their invaluable help.
We also thank Dr. Hayashino for his invaluable help.  
This work was financially supported in part by
the Ministry of Education, Calture, Sports, Science, and Technology 
(Nos. 10044052 and 10304013).


\vspace{1cm}

\begin{figure}
\begin{center}
\FigureFile(120mm,120mm){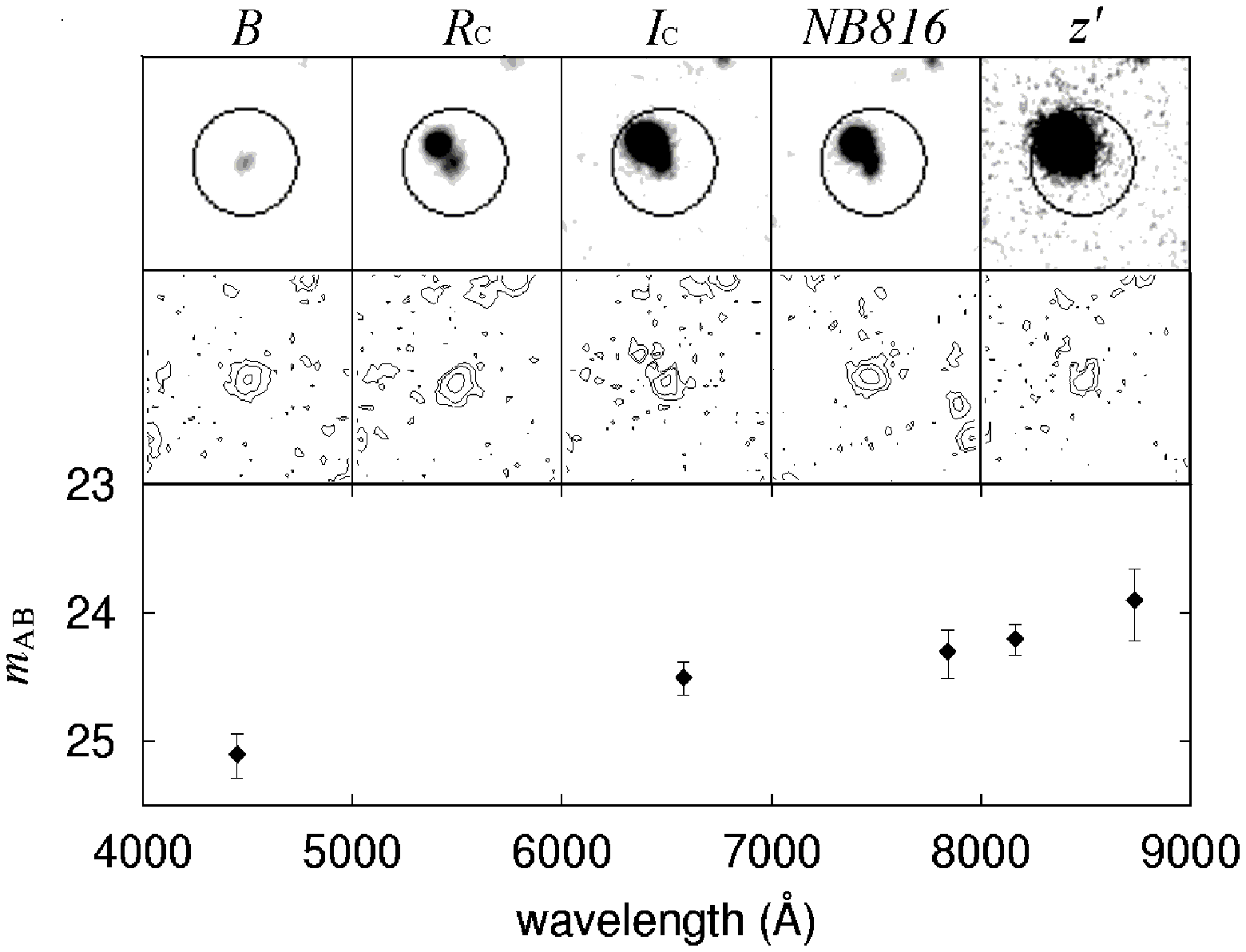}
\end{center}
\caption{Thumb-nail images of  SDSSp J104433.04-012502.2
and its neighbor galaxy (upper panel).
The angular size of the circle in each panel corresponds to
$8^{\prime \prime}$.
A contour plot of a lens galaxy after the quasar had been subtracted 
is shown in the middle panel. The lower panel shows
the spectral energy distribution of the neighbor galaxy
in the magnitude scale.}
\label{fig:fig1}
\end{figure}

\begin{figure}
\begin{center}
\FigureFile(120mm,120mm){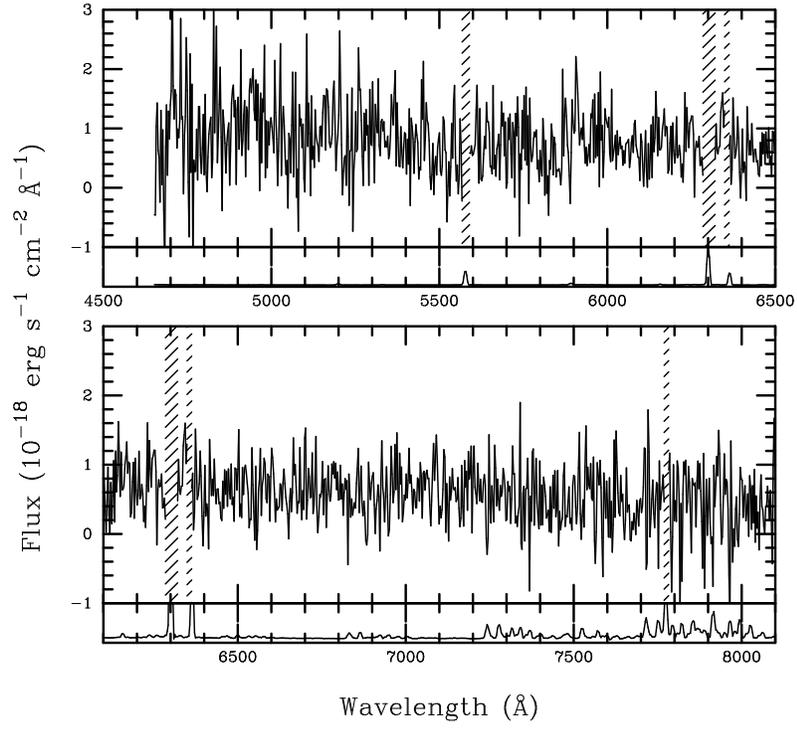}
\end{center}
\caption{Optical spectrum of the neighboring galaxy.
The hatched regions indicate the positions of the strong night sky lines.  
The sky spectrum is also shown in the lower part of each panel.
}
\label{fig:fig2}
\end{figure}

\begin{figure}
\begin{center}
\FigureFile(120mm,120mm){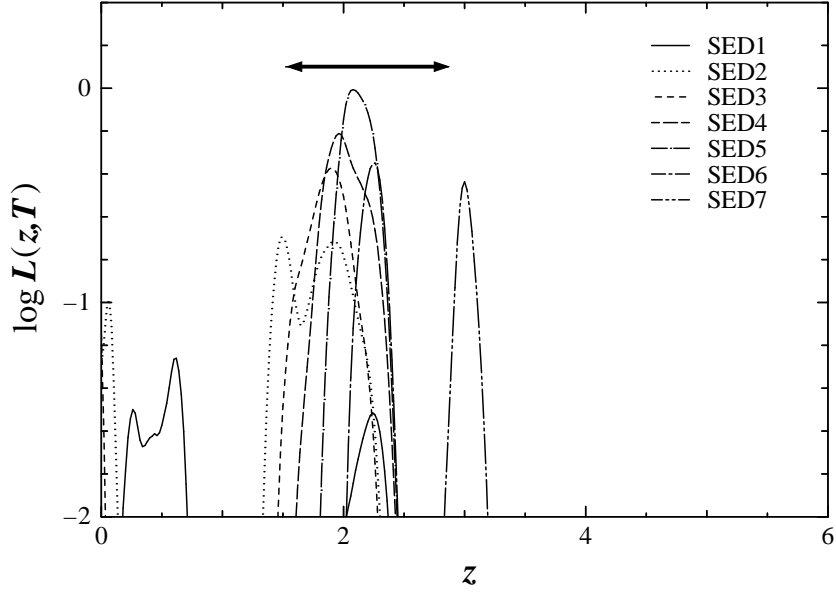}
\end{center}
\caption{Distributions of the likelihood for the seven SEDs
shown as  a function of the redshift.
The arrow in the upper part shows the range of the probable redshift of 
the galaxy estimated by optical spectroscopy.}
\label{fig:fig3}
\end{figure}

\begin{figure}
\begin{center}
\FigureFile(120mm,120mm){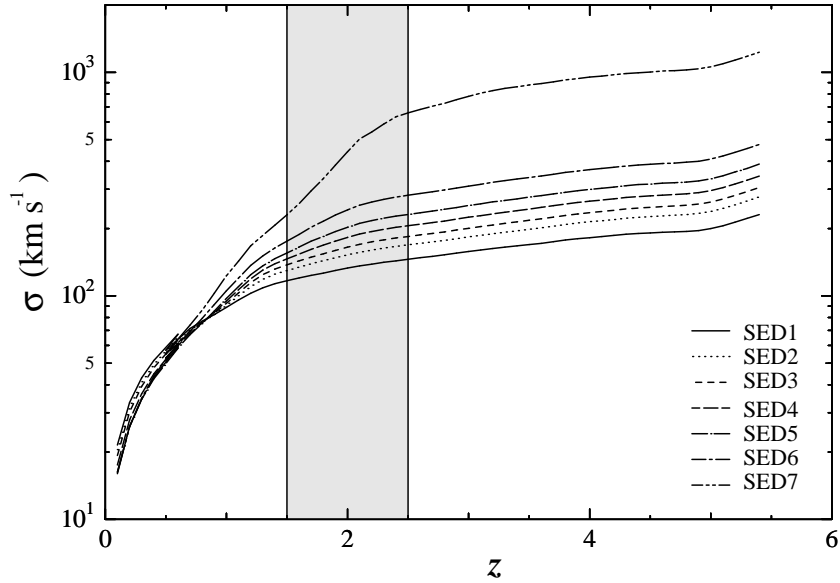}
\end{center}
\caption{Distributions of the velocity dispersion 
of the lens galaxy shown as a function of the redshift for seven SEDs.}
\label{fig:fig4}
\end{figure}

\begin{figure}
\begin{center}
\FigureFile(100mm,100mm){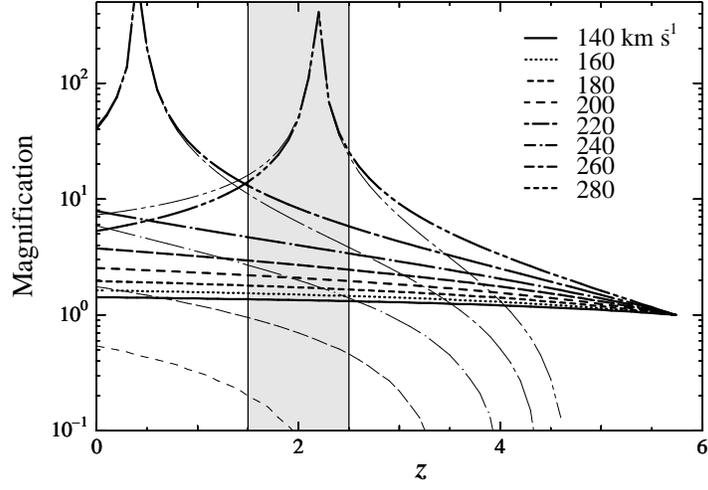}
\end{center}
\caption{Distributions of the magnification factor of 
the source observed at $\theta = \timeform{1".9}$  (thick lines) and 
its counter image (thin lines) shown 
as a function of the redshift for the eight cases for the stellar 
velocity dispersion.
We assume the magnification of the former one to be $M_+$, 
except for the case that $\theta$ is smaller than $\theta_{\rm E}$.}
\label{fig:fig5}
\end{figure}

\clearpage
\begin{table}
\caption{List of $z \sim$ 6 SDSS quasars.}\label{tab:tab1}
\begin{center}
\begin{tabular}{cccc}
\hline
\hline
Name &
Redshift &
$z^\prime$  &
$M_B$$^*$ \\
\hline
SDSSp J104433.04$-$012502.2 & 5.74$^{\dag}$ & 19.20 & $-$28.09 \\
SDSSp J083643.85+005453.3   & 5.82 & 18.74 & $-$28.54 \\
SDSSp J130608.26+035626.3   & 5.99 & 19.47 & $-$27.73 \\
SDSSp J103027.10+052455.0   & 6.28 & 20.05 & $-$27.23 \\
\hline
\end{tabular}
\end{center}

$^*$The $k$-correction was made by using the composite
quasar spectrum in Vanden Berk et al. (2001).\\
$^{\dag}$The discovery redshift was $z=5.8$ (Fan et al. 2000).
Since, however, the subsequent optical spectroscopic observations
suggested a bit lower redshift; $z=5.73$ (Djorgovski et al. 2001)
and $z=5.745$ (Goodrich et al. 2001), we adopt $z=5.74$ in this paper
 for quasar spectrum in Vanden Berk et al. (2001).
\end{table}

\begin{table}
\caption{Journal of imaging observations.}\label{tab:tab2}
\begin{center}
\begin{tabular}{lcccc}
\hline
\hline
Band & 
Obs. date (UT) & 
Total integ. time (s)  &
$m_{\rm lim}$(AB)$^*$ &
FWHM$_{\rm star}^{\dag}$ ($^{\prime \prime}$) \\
\hline
$B$         & 2002 February 17      &  1680 & 27.1 & 1.2 \\
$R_{\rm C}$ & 2002 February 15, 16  &  4800 & 26.8 & 1.4 \\
$I_{\rm C}$ & 2002 February 15, 16  &  3360 & 26.2 & 1.2 \\
NB816       & 2002 February 15 --- 17 & 36000 & 26.6 & 0.9 \\
$z'$        & 2002 February 15, 16  &  5160 & 25.4 & 1.2 \\
\hline
\end{tabular}
\end{center}

$^*$ The limiting magnitude (3$\sigma$) with a
2$^{\prime\prime}$ aperture.\\
$^{\dag}$ The full width at half maximum of stellar
objects in the final image.\\
\end{table}

\begin{table}
\caption{Photometry of a lens galaxy.}\label{tab:tab3}
\begin{center}
\begin{tabular}{lccccc}
\hline
\hline
Object & $B$ & $R_{\rm C}$ & $I_{\rm C}$ & NB816 & $z^{\prime}$ \\
\hline
Lens galaxy & $25.1 \pm 0.06$ & $24.5 \pm 0.04$ & $24.3 \pm 0.06$ & $24.2 \pm 0.04$ & $23.9 \pm 0.09$ \\
\hline
\end{tabular}
\end{center}
\end{table}


\begin{thebibliography}{}
\bibitem[]{}Ajiki, M., Taniguchi, Y., Murayama, T., Nagao, T., Veilleux,
	  S., Shioya, Y., Fujita, S. S., Kakazu, Y. 2002, ApJ, 576, L25
\bibitem[]{}Barkana, R., \& Loeb, A. 2000, ApJ, 531, 613
\bibitem[]{}Binney, J., \& Merrifield, M. 1998, Galactic Astronomy,
          (Princeton, New Jersey, Princeton University Press), Ch. 2.4
\bibitem[]{}Bruzual, A. G., \& Charlot, S. 1993, ApJ, 405, 538
\bibitem[]{}Calzetti, D., Armus, L.,  Bohlin, R. C., Kinney, A. L.,
              Koornneef, J.,  \& Storchi-Bergmann, T. 2000, ApJ, 533, 682
\bibitem[]{}Coleman, G. D., Wu, C.-C., \& Weedman, D. W. 1983, ApJS, 43, 393
\bibitem[]{}Djorgovski, S. G., Castro, S., Stern, D., \& Mahabal,
              A. A. 2001, ApJ, 560, L5
\bibitem[]{}Fan, X., Narayanan, V. K., Lupton, R. H., Strauss, M. A.,
	  Knapp, G. R., Becker, R. H., White, R. L., Pentericci, L., et
	  al. 2001, AJ, 122, 2833
\bibitem[]{}Fan, X., White, R. L., Davis, M., Becker, R. H., Strauss, M.
	  A., Haiman, Z., Schneider, D. P., Gregg, M. D., et al. 2000, AJ, 120, 1167
\bibitem[]{}Goodrich, R. W., Campbell, R., Chaffee, F. H., Hill, G. M.,
	  Sprayberry, D., Brandt, W. N., Schneider, D. P., Kaspi, S., et
	  al. 2001, ApJ, 561, L23
\bibitem[]{}Haiman, Z., \& Loeb, A. 2001, 503, 505
\bibitem[]{}Kaifu, N. 1998, Proc. SPIE, 3352, 14
\bibitem[]{}Kashikawa, N., Inata, M., Iye, M., Kawabata, K., Okita, K.,
	  Kosugi, G., Ohyama, Y., Sasaki, T. et al. 2000, Proc. SPIE, 4008, 104
\bibitem[]{}Leitherer, C., Alloin, D., Fritze-von Alvensleben, U.,
	  Gallagher, J. S., Huchra, J. P., Matteucci, F., O'Connell,
	  R. W., Beckman, J. E., et al. 1996, PASP, 108, 966
\bibitem[]{}Madau, P., Ferguson, H. C., Dickinson, M. E., Giavalisco, M.,
              Steidel, C. C., \& Fruchter, A. 1996, MNRAS, 283, 1388
\bibitem[]{}Massey, P., Strobel, K., Barnes, J. V., \& Anderson, E. 1988,
              ApJ, 328, 315
\bibitem[]{}Miyazaki, S., Sekiguchi, M., Imi, K., Okada, N., Nakata, F., \& 
              Komiyama, Y.\ 1998, Proc. SPIE, 3355, 363
\bibitem[]{}Oke, J. B. 1990, AJ, 99, 1621
\bibitem[]{}Sakai, S., Mould, J. R., Hughes, S. M. G., Huchra, J. P.,
	  Macri, L. M., Kennicutt, R. C., Jr., Gibson, B. K., Ferrarese,
	  L., et al. 2000, ApJ, 529, 698
\bibitem[]{}Scott, J., Bechtold, J., \& Dobrzycki, A.
              2000, ApJS, 130, 37
\bibitem[]{}Turner, E. L. 1991, AJ, 101, 5
\bibitem[]{}Vanden Berk, D. E., Richards, G. T., Bauer, A., Strauss,
	  M. A., Schneider, D. P., Heckman, T. M., York, D. G., Hall,
	  P. B., et al. 2001, AJ, 122, 549
\bibitem[]{}Wyithe, J. S. B., \& Loeb, A. 2002, Nature, 417, 923 (WL02)
\bibitem[]{}Yagi M.,  Kashikawa, N., Sekiguchi, M., Doi, M., Yasuda, N.,
              Shimasaku, K., \& Okamura, S. 2002, AJ, 123, 66
\bibitem[]{}York, D. G., Adelman, J., Anderson, J. E., Jr., Anderson,
	  S. F., Annis, J., Bahcall, N. A., Bakken, J. A., Barkhauser,
	  R., et al. 2000, AJ, 120, 1579
\end{thebibliography}
\end{document}